\documentclass[amsmath, amsfonts, amssymb, aps, showkeys, twocolumn]{revtex4-1}

\usepackage[utf8]{inputenc}
\usepackage[T1]{fontenc}
\usepackage{graphicx}
\usepackage{dcolumn}
\usepackage{bm}
\usepackage{tikz}
\usepackage{censor}
\usepackage{url}

\begin{document}


\title{A Ballistic Monte Carlo Approximation of $\pi$}

\author{Vincent Dumoulin}
\email{vincent.dumoulin@umontreal.ca}
\affiliation{
    Département d'informatique et de recherche opérationnelle \\
    Université de Montréal
}
\author{Félix Thouin}
\email{felix.thouin@umontreal.ca}
\affiliation{
    Département de physique \\
    Université de Montréal
}

\date{\today}

\begin{abstract}
We compute a Monte Carlo approximation of $\pi$ using importance sampling with
shots coming out of a Mossberg 500 pump-action shotgun as the proposal
distribution. An approximated value of $3.131$ is obtained, corresponding to a
$0.33\%$ error on the exact value of $\pi$. To our knowledge, this represents
the first attempt at estimating $\pi$ using such method, thus opening up
new perspectives towards computing mathematical constants using everyday tools.
\end{abstract}

\maketitle


\section{Introduction}

The ratio between a circle's circumference and its diameter, named $\pi$, is a
mathematical constant of crucial importance to science, yet most scientists rely
on pre-computed approximations of $\pi$ for their research. This is problematic,
because scientific progress relies on information that will very likely
disappear in case of a cataclysmic event, such as a zombie apocalypse. In such
case, scientific progress might even stop entirely. This motivates the need for
a robust, yet easily applicable method to estimate $\pi$.

We first lay down the theoretical framework for Monte Carlo methods, including
importance sampling and propose a probabilistic interpretation of $\pi$
within this framework. We then introduce the idea of computing $\pi$ using
importance sampling with a ballistic-based proposal distribution and suggest a
robust way of dealing with the unknown generating distribution. Finally, we
compare the obtained estimation of $\pi$ with the true value.

\subsection{Monte Carlo}

The Monte Carlo method, first introduced in \cite{metropolis1949monte}, is a
stochastic approach to computing expectations of functions of random variables.
Let $f(\mathbf{x}): \mathbb{R}^n \rightarrow \mathbb{R}^+$ be a probability
density function over a random vector $\mathbf{x} \in \mathbb{R}^n$ and let
$g(\mathbf{x}): \mathbb{R}^n \rightarrow \mathbb{R}$ be a function of $\mathbf{x}$.

The expected value $\mathbb{E}$ of $g$ over $f$ is defined as

\begin{equation}
\label{eqn:expectation}
    \mathbb{E}_f [g(\mathbf{x})]
    = \int_{\mathbb{R}^n} f(\mathbf{x}) g(\mathbf{x}) d\mathbf{x}.
\end{equation}

The Monte Carlo method approaches $\mathbb{E}_f [g(\mathbf{x})]$ with

\begin{equation}
    \mathbb{E}_f [g(\mathbf{x})]
    \approx \hat{\mathbb{E}}_f [g(\mathbf{x})]
    = \frac{1}{N} \sum_{i=1}^N g(\mathbf{x}_i), \quad
    \mathbf{x}_i \sim f(\mathbf{x})
\end{equation}
where $\mathbf{x}_i \sim f(\mathbf{x})$ means $\mathbf{x}_i$ is drawn
$f(\mathbf{x})$.

Note that $\hat{\mathbb{E}}_f [g(\mathbf{x})]$ is a consistent estimator of
$\mathbb{E}_f [g(\mathbf{x})]$, i.e. $\hat{\mathbb{E}}_f [g(\mathbf{x})]$
converges in probability to $\mathbb{E}_f [g(\mathbf{x})]$ as $N \rightarrow
\infty$. Furthermore, its variance decreases as $\frac{1}{N}$
\emph{independently of the dimensionality of $\mathbf{x}$}. For more details,
see \cite{bishop2006pattern}.

\subsection{Importance sampling}

When sampling from $f(\mathbf{x})$ is difficult or impossible, or when
$f(\mathbf{x})$ is too different from $g(\mathbf{x})$ (i.e. high probability
mass regions correspond to a low-valued $g(\mathbf{x})$ and vice-versa), Monte
Carlo methods may fail. In that case, we note that

\begin{equation}
\begin{split}
    \mathbb{E}_f [g(\mathbf{x})]
    &= \int_{\mathbb{R}^n} f(\mathbf{x}) g(\mathbf{x}) d\mathbf{x} \\
    &= \int_{\mathbb{R}^n} q(\mathbf{x})\frac{f(\mathbf{x})}{q(\mathbf{x})}
            g(\mathbf{x})d\mathbf{x} \\
    &= \mathbb{E}_q\left[\frac{f(\mathbf{x})}{q(\mathbf{x})}g(\mathbf{x})\right] \\
\end{split}
\end{equation}
for some arbitrary distribution $q(\mathbf{x})$ such that $q(\mathbf{x}) > 0$
when $f(\mathbf{x})g(\mathbf{x}) \neq 0$. We can therefore reformulate the
approximation as

\begin{equation}
\begin{split}
    \mathbb{E}_f [g(\mathbf{x})]
    &\approx \hat{\mathbb{E}}_q
             \left[\frac{f(\mathbf{x})}{q(\mathbf{x})}g(\mathbf{x})\right]\\
    &= \frac{1}{N} \sum_{i=1}^N
       \frac{f(\mathbf{x}_i)}{q(\mathbf{x}_i)}g(\mathbf{x}_i), \quad
    \mathbf{x}_i \sim q(\mathbf{x}).
\end{split}
\end{equation}

This method is called \emph{importance sampling}, and given a careful choice of
$q(\mathbf{x})$, can allow easier sampling and help reduce the variance of the
estimator. We note once again that $\hat{\mathbb{E}}_q \left[\frac{f(\mathbf{x})}
{q(\mathbf{x})}g(\mathbf{x})\right]$ is a coherent estimator of
$\mathbb{E}_f [g(\mathbf{x})]$. For more details, see \cite{bishop2006pattern}.

\subsection{Introducting $\pi$ in the Monte Carlo framework}

\begin{figure}[t]
\centering
\begin{tikzpicture}
    \filldraw[fill=black!40!white, draw=black!40!white] (-0.85,-0.85) --
    (4,-0.85) -- (-0.85,4);
    \filldraw[fill=black!40!white, draw=black!40!white] (4,-0.85) arc (0:90:4.85);
    \draw (4,-0.85) arc (0:90:4.85);
    \draw (-0.85,-0.85) -- (4,-0.85) -- (4,4) -- (-0.85,4) -- (-0.85,-0.85);
    \draw (-1.5,1.575) node {1};
\end{tikzpicture}
\caption{\label{fig:square_circle} $\pi$ is proportional to the area of a unit
         square occupied by a quarter circle of radius $1$.}
\end{figure}
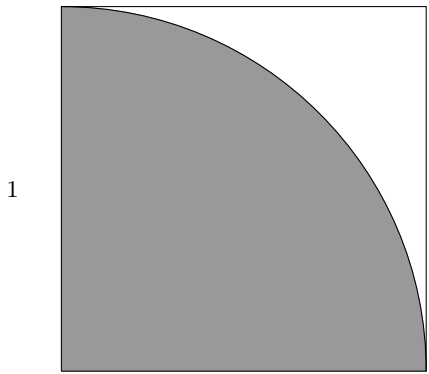

By choosing an appropriate probability density function $f(\mathbf{x})$ and an
appropriate function $g(\mathbf{x})$, the value of $\pi$ can be approached using
a Monte Carlo estimation.

Consider a unit square and a circle arc joining two opposite corners of the
square (Fig.~\ref{fig:square_circle}). The area of the square is $1$, while the
area of the quarter circle is $\frac{\pi}{4}$, which means the proportion of the
square's area occupied by the quarter circle is also $\frac{\pi}{4}$. In other
words,

\begin{equation}
\label{eqn:circle_area}
    \int_{0}^1\int_{0}^1 g(x, y) dx dy = \frac{\pi}{4}
\end{equation}
with
\begin{equation}
    g(x, y) = 
    \begin{cases}
        1, \quad x^2 + y^2 \leq 1, \\
        0, \quad \text{otherwise}.
    \end{cases}
\end{equation}

We rewrite equation \ref{eqn:circle_area} as
\begin{equation}
\label{eqn:extended_circle_area}
    \int_{-\infty}^\infty\int_{-\infty}^\infty
        f(x, y)g(x, y) dx dy = \frac{\pi}{4}
\end{equation}
with
\begin{equation}
    f(x, y) = 
    \begin{cases}
        1, \quad 0 \leq x \leq 1 \text{ and } 0 \leq y \leq 1, \\
        0, \quad \text{otherwise}
    \end{cases}
\end{equation}
and observe that it is identical in form to equation \ref{eqn:expectation}.
Therefore the numerical value of $\pi$ can be interpreted as
the expected value of $g(\mathbf{x})$ with $\mathbf{x}$ being a random
two-dimensional vector uniformly distributed over the unit square.

This allows us to approximate equation \ref{eqn:extended_circle_area} as

\begin{equation}
\begin{split}
    \frac{\pi}{4}
    \approx \hat{\mathbb{E}}_f [g(\mathbf{x})]
    &= \frac{1}{N} \sum_{i=1}^N g(\mathbf{x}_i), \\
    \mathbf{x}_i \sim f(\mathbf{x}) &= \mathcal{U}([0 , 1] \times [0, 1])
\end{split}
\end{equation}

In other words, to approximate $\frac{\pi}{4}$, one needs to draw $N$
uniformly-distributed samples across the unit square and count the proportion
of those points which fall into the quarter circle.

\subsection{Off-the-shelf random sampling}

\begin{figure}[t]
\centering
\begin{tikzpicture}
    \draw (2,4) -- (4,3) -- (4,0) -- (2,1) -- (2,4);
    \draw (5.2,1.4) node {Aluminum foil};
    \draw (0,0) -- (3,3);
    \draw (3,3) node {.};
    \draw (0,0) -- (3,2.5);
    \draw (3,2.5) node {.};
    \draw (0,0) -- (2.4,2.9);
    \draw (2.4,2.9) node {.};
    \draw (0,0) -- (3.8,2.5);
    \draw (3.8,2.5) node {.};
    \draw (0,0) -- (2.8,1.5);
    \draw (2.8,1.5) node {.};
    \draw (0,0) -- (3.5,1.5);
    \draw (3.5,1.5) node {.};
    \draw (-1.2,-0.65) -- (-0.20,0.0);
    \draw (-1,-0.85) -- (0,-0.20);
    \draw (0,-0.20) arc (0:90:0.2);
    \draw (1,-0.20) node {Shotgun};
\end{tikzpicture}
\caption{\label{fig:setup} Experimental setup. The shotgun is pointed at an
         aluminum foil to record its shot pattern. Samples are iteratively drawn
         by shooting the shotgun.}
\end{figure}
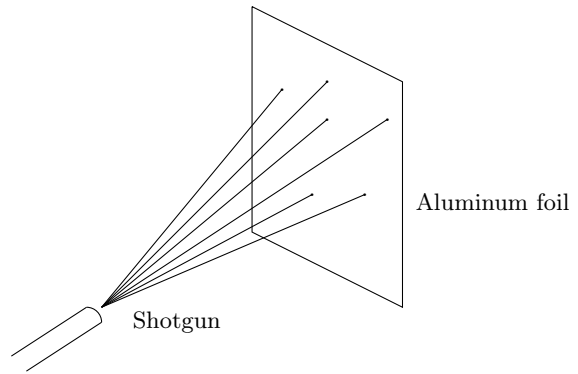

In order to estimate $\pi$ using a Monte Carlo method, one needs to draw
independent and identically-distributed (IID) samples from a uniform
distribution. While computer-assisted pseudo-random number generation is
computationally cheap and fast, it relies on technology which might not be
available in the event of a zombie apocalypse.

On the other hand, primitive methods such as coin tosses or dice throws are
almost always readily available, but they are slow and their sampling time scales
linearly with the numerical precision required.

With this in mind, we advocate the use of ballistic-assisted (i.e.
projectile-based) random sampling methods because they are both easily
accessible and parallelizable. In particular, shotgun-assisted random sampling
seems very suitable because of the presumed abundance of shotguns in
cataclysmic times and the speed at which they can generate samples.

To our knowledge, no prior attempt at estimating $\pi$ using ballistic-assisted
random sampling methods has ever been made.

\subsection{Uniformity issue}

There is no guarantee that shotgun shots are uniformly distributed. In fact,
pellet distribution depends on many latent variables such as height of the
shooter, distance to the target, orientation of the shotgun and wind direction,
to name a few. The issue can be overcome by using importance sampling, but the
shot distribution is unknown, and will need to be estimated in order for
importance sampling to work.

Since this instance of density estimation problem is low-dimensional, a simple
histogram method is sufficient. The bin width hyperparameter can be decided using
cross-validation on the log-likelihood of the samples.

Thus $\frac{\pi}{4}$ can be computed as
\begin{equation}
\label{eqn:experiment_eqn}
\begin{split}
    \frac{\pi}{4}
    \approx \hat{\mathbb{E}}_f \left[\frac{f(\mathbf{x})}{\hat{q}(\mathbf{x})}g(\mathbf{x})\right]
    &= \frac{1}{N} \sum_{i=1}^N \frac{g(\mathbf{x}_i)}{\hat{q}(\mathbf{x}_i)}, \\
    \mathbf{x}_i \sim q(\mathbf{x})
\end{split}
\end{equation}
where $\hat{q}(\mathbf{x})$ is the histogram estimate of the probability density
function (PDF).

\section{Experimental setup}
\label{sec:setup}

\begin{figure}[t]
    \includegraphics[width=\columnwidth]{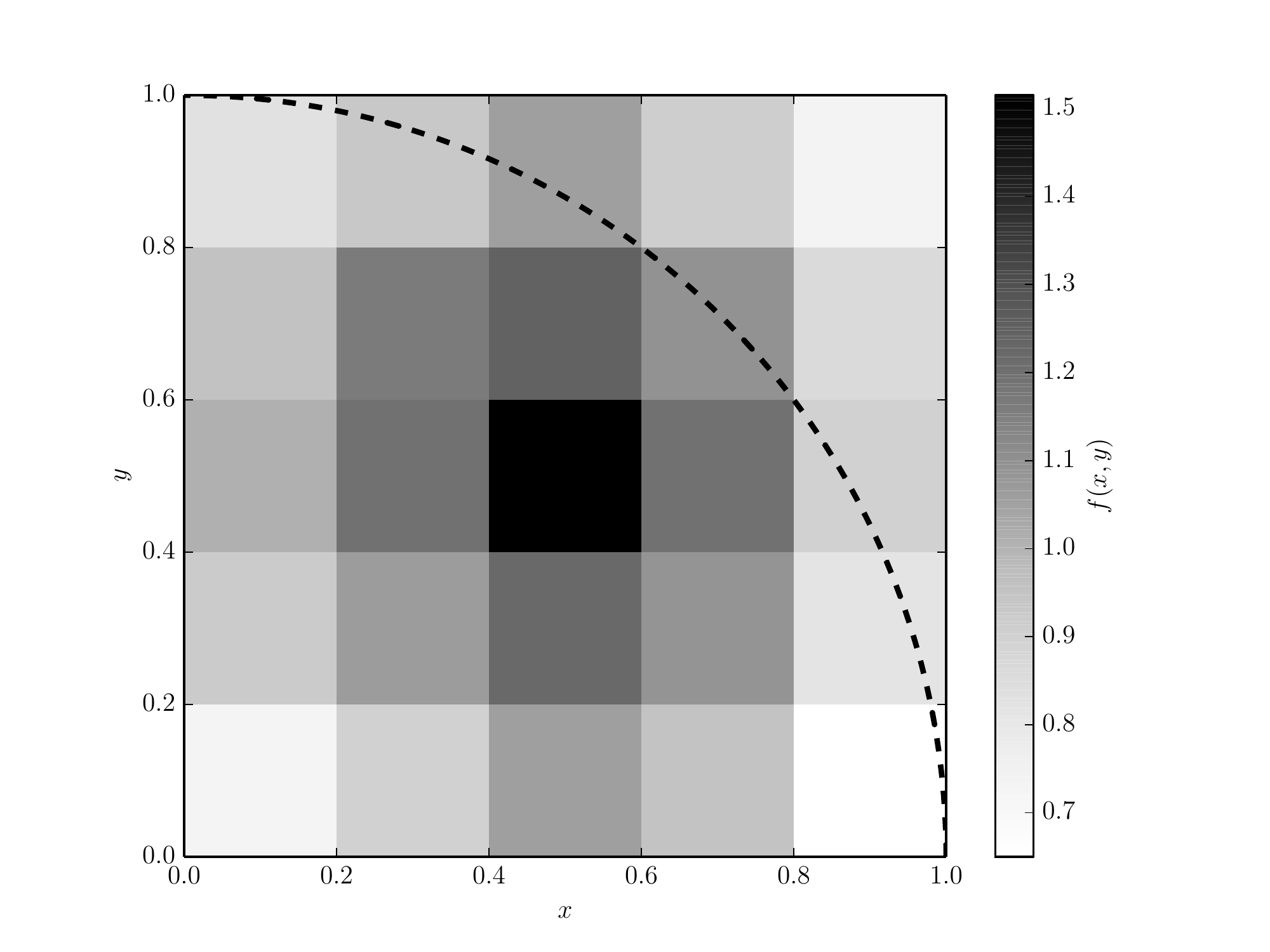}
    \caption{\label{fig:pdf_estimation} An estimate of the probability density
             function of the data points obtained using a 2D histogram method
             with a bin width of $0.2$ on $10000$ randomly-selected samples. A
             quarter circle was drawn for visual reference.}
\end{figure}

To sample from the proposal distribution, a 28-inch-barrel Mossberg 500 shotgun
was used. The latter is chambered for 3 inches, 12-gauge shotshells and its spread
can be tuned by using different types of chokes mounted at the end of the
barrel.

For this experiment, we used an improved cylinder choke made by Mossberg.
Cartriges used were composed of 3 dram equivalent of powder and 32 grams of \#8
lead pellets. Average muzzle velocity is estimated to be around $366 m/s$ by the
manufacturer.

Samples were recorded by placing an aluminum foil in the trajectory of
the shots (Fig.~\ref{fig:setup}). The shotgun was fired at a $20m$ distance from
the targets.

Foils were photographed and samples were extracted by locating holes in the image
and computing their center of mass. Sample positions were normalized to be in
$[0, 1] \times [0, 1]$
\footnote{Code and data points used for this experiment are available at
\protect\url{https://github.com/vdumoulin/research/tree/master/code/shotgun_monte_carlo}}.

\section{Results}

\begin{figure}[t]
    \includegraphics[width=\columnwidth]{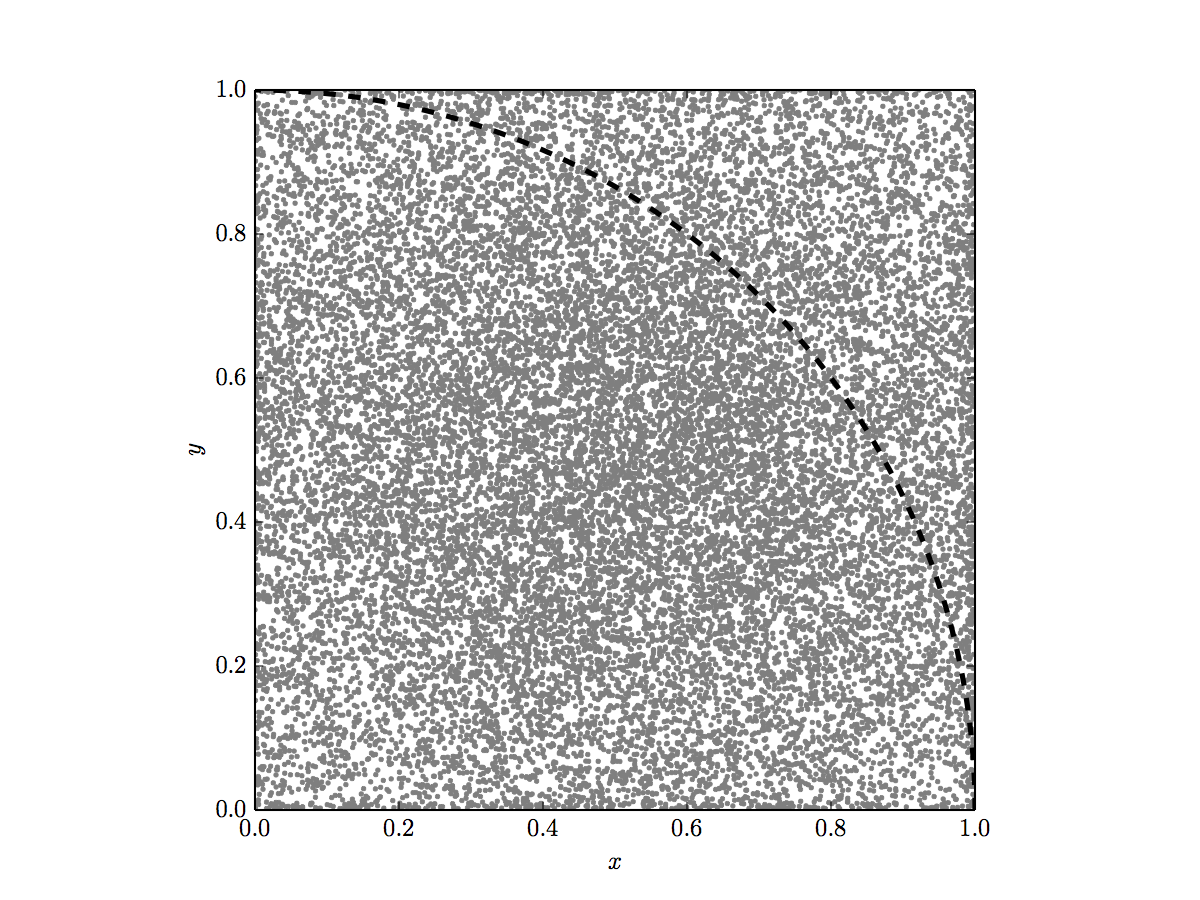}
    \caption{\label{fig:data_points} Remaining $20857$ samples used to
             approximate $\pi$. Positions were normalized to be in
             $[0, 1] \times [0, 1]$. A quarter circle was
             drawn for visual reference.}
\end{figure}

The experiment described above was carried by firing 200 shots, yielding $30857$
samples.

The sample extraction method described in section \ref{sec:setup} could not
distinguish holes made by one or many pellets.  One sample might actually
correspond to many pellet impacts, or even a wad impact, but this factor of
variation is integrated in the PDF estimation.

Of the $30857$ samples produced, we used a random subset of $10000$ samples for
PDF estimation. The optimal bin width hyperparameter was determined to be $0.2$
using 20-fold cross-validation. The estimated PDF of the shot
distribution is shown in Fig.~\ref{fig:pdf_estimation}.

Using equation~\ref{eqn:experiment_eqn} on the $20857$ remaining samples
(Fig.~\ref{fig:data_points}), a value of $3.131$ for $\pi$ was obtained, which
corresponds to a $0.33\%$ error on the true value.

\section{Conclusion}

We successfully estimated the value of $\pi$ using a Monte Carlo method by
drawing samples from shots coming out of a Mossberg 500 pump-action shotgun.
Non-uniformity of pellet spread was accounted for by importance sampling.
The pellet distribution was approximated using a 2D histograms method.

While variance on the estimate could be reduced by drawing more samples, this
is still a striking display of the robustness of Monte Carlo methods:
even though pellet distribution depended on many uncontrolled factors (wind
direction, muzzle orientation, aluminium foil geometry, and wad impacts to name
a few), the approached value of $\pi$ ($3.131$) is still within $0.33\%$ of the
true value.

We feel confident that ballistic Monte Carlo methods such as the one presented in
this paper constitute reliable ways of computing mathematical constants should a
tremendous civilization collapse occur.

\begin{acknowledgments} 
We would like to thank Geoffroy Bergeron for providing a safe environment to
perform the experiment and Marie-Hélène Labrecque for logistic support. We
would also like to thank Olivier Mastropietro for his help in carrying the
experiment.
\end{acknowledgments} 

\bibliography{references}
\end{document}